\documentclass{jpsj3}  

\title{Flux Phase in Bilayer $t-J$ Model
 }

\author{\name{Kazuhiro \surname{Kuboki}}\thanks{E-mail address: kuboki@kobe-u.ac.jp} 
}
\inst{\address{Department of Physics, Kobe University, 
Kobe 657-8501, Japan} 
}

\begin{document}
\maketitle

Various experiments suggest that time-reversal symmetry (${\cal T}$) 
is  broken spontaneously in some of the high-$T_C$ cuprate
superconductors.\cite{Coving,TRSB1,TRSB2,TRSB3}
For example, Covington {\it et al.}\cite{Coving} observed the peak splitting 
of zero bias conductance in ab-oriented YBCO/insulator/Cu junctions.
This has been interpreted as a sign of ${\cal T}$ violation
caused by the introduction of superconducting (SC) order 
parameter (OP) with a symmetry different from that in the 
bulk.\cite{Fogel,Matsu1,Matsu2} 
In this case, spontaneous currents would flow along the surface,  
and a magnetic field should be generated locally.
However, experimental evidence for such magnetic fields 
is still controversial.\cite{Carmi,msr}

Recently, the present author has studied the (110) surface state of high-$T_C$ 
cuprate superconductors based on the Bogoliubov de Gennes (BdG) method
applied to a single-layer $t-J$ model, and 
it was found that the flux phase can occur as a surface state.\cite{KK1,KK2} 
The flux phase is a mean-field  (MF) solution to the 
$t-J$ model in which staggered currents flow and the flux penetrates the plaquette 
in a square lattice, but it is unstable toward the $d_{x^2-y^2}$-wave SC 
instability.\cite{Affleck,Zhang1,Hamada,Bejas,Zhao} 
(The $d$-density wave states, which have been introduced in a
different context, have similar properties.\cite{Chakra})
Near the (110) surfaces, the  $d_{x^2-y^2}$-wave SC order is strongly 
suppressed and then the flux phase that is forbidden in the bulk may arise.\cite{KK2} 
Once it occurs, the spontaneous currents flow along the surface, 
leading to local ${\cal T}$ violation. 
However, the doping range in which ${\cal T}$ violation arises 
was much narrower than that observed experimentally in YBCO, 
if we use an effective single-layer model.\cite{tana}

In this short note, we study the bare transition temperature of the 
flux phase (assuming the absence of SC order), $T_{FL}$, 
in a bilayer $t-J$ model that describes the electronic 
states of the YBCO system more accurately.
The critical doping rate $\delta_c$, at which $T_{FL}$ vanishes, is 
estimated and compared with that in the single-layer model. 
In bilayer models, there may be two types of flux phase, {\it i.e.}, 
the directions of the flux in two layers are the same or opposite. 
When the latter state arises, the magnetic fields generated in two layers cancel 
each other.

We consider the bilayer $t-J$ model on a square lattice whose Hamiltonian is 
given by $H  =  H_1 + H_2 + H_\perp$ with  
\begin{eqnarray}
\displaystyle 
H_i =&\displaystyle -\sum_{j,\ell,\sigma} 
t_{j\ell} {\tilde c}^{(i)\dagger}_{j\sigma} {\tilde c}^{(i)}_{\ell\sigma}
 +J\sum_{\langle j,\ell\rangle} {\bf S}^{(i)}_j\cdot {\bf S}^{(i)}_\ell, 
 \ \ \  (i=1,2)\\
 H_\perp =&\displaystyle -\sum_{j,\ell,\sigma} t_{j\ell}^\perp 
\Big( {\tilde c}^{(1)\dagger}_{j\sigma} {\tilde c}^{(2)}_{l\sigma} + h.c.\Big)
 +J_\perp\sum_j {\bf S}^{(1)}_j\cdot {\bf S}^{(2)}_j, 
\end{eqnarray}
where the transfer integrals (in plane) $t_{j\ell}$ are finite for the first-  ($t$), 
second-  ($t'$), and third-nearest-neighbor bonds ($t''$), or zero otherwise.  
$J$ $(J_\perp)$ is the intraplane (interplane) antiferromagnetic superexchange interaction, and $\langle j,\ell \rangle$ denotes nearest-neighbor bonds. 
The interplane transfer integrals $t^\perp_{j\ell}$ are chosen to reproduce 
the dispersion in $k$ space,\cite{Andersen} 
$t^\perp_k =  -t^\perp_0 (\cos k_x - \cos k_y)^2$, namely, 
"on-site" ($t^\perp_0$), second- ($t^\perp_2 = -t^\perp_0/2$) , 
and third-nearest-nearest-neighbor bonds ($t^\perp_3 = t^\perp_0/4$) 
are taken into account. 

${\tilde c}^{(i)}_{j\sigma}$ is the electron operator for the $i$-th layer 
($i=1,2$) in Fock space without double occupancy. 
We treat this condition using the slave-boson MF 
theory.\cite{Zou,Ogata,Lee,Hamada,KK1,KK2} 
Although the bosons are not condensed in purely two-dimensional 
systems  at a finite temperature ($T$), they are almost condensed at a low $T$ 
({\it i.e.}, $T < 3J/16$ where the flux phase may occur) and for 
finite carrier doping ($\delta \gtrsim 0.05$). 
Then, we treat them as Bose-condensed.
(For a small $\delta$, the absence of Bose condensation may lead to 
a flux phase as a stable solution.\cite{Hamada,Zhao})
This procedure amounts to renormalizing  the transfer integrals 
by multiplying $\delta$ ($\delta$ being the doping rate), 
{\it e.g.}, $t \to t\delta$, {\it etc.}, 
and rewriting ${\tilde c}_{j\sigma}$ as $f_{j\sigma}$. 
In a qualitative sense, this approach is equivalent to 
the renormalized mean-field theory of Zhang {\it et al.}\cite{Zhang2} 
(Gutzwiller approximation). 

We decouple the Hamiltonian by dividing the system 
into two sublattices A and B. 
The bond OPs may be  complex numbers 
when the flux order occurs, and we define intralayer OPs as  
$ \sum_\sigma \langle f^{(i)\dagger}_{j\sigma}f^{(i)}_{j+{\hat x}\sigma} \rangle 
\equiv x_s + i(-1)^{j_x+j_y}y_s$, 
$ \sum_\sigma \langle f^{(i)\dagger}_{j\sigma}f^{(i)}_{j+{\hat y}\sigma} \rangle 
\equiv x_s -  i(-1)^{j_x+j_y}y_s$. 
Here, ${\hat x}$ (${\hat y}$) is a unit vector in the $x$ 
($y$)-direction (the lattice constant is taken to be unity), 
and $x_s$ and $y_s$ are real constants. 
For interlayer bonds, we define 
$\sum_\sigma \langle f^{(1)\dagger}_{j\sigma} f^{(2)}_{j\sigma}\rangle 
\equiv x_s^\perp$, with $x_s^\perp$ being a real constant.
Now we note that there are two ways of coupling the layers;  
a site in the A sublattice in one layer may be on top of a site in the A (or B) sublattice 
of the other layer. 
We call the former (latter) one as a type A (B) flux phase. 

Energy eigenvalues are obtained by diagonalizing the Hamiltonian.  
They are given as  
$
E^{(\alpha\beta)}_k =  \alpha |\xi_k| + \beta\epsilon^\perp_k + \epsilon_k  
$
for the type A,   
$
E^{(\alpha\beta)}_k = \alpha |\xi_k + \beta \epsilon^\perp_k| + \epsilon_k  
$
for the type B flux phase, respectively, with $\alpha, \beta = \pm$. 
Here, 
${\rm Re} \ \xi_k =   -(2t\delta +3Jx_s/4) (\cos k_x + \cos k_y)$, 
${\rm Im} \ \xi_k = -3Jy_s /4 (\cos k_x - \cos k_y)$,   
$\epsilon_k  =  -\mu -4t'\delta \cos k_x \cos k_y 
-2t''\delta (\cos 2k_x + \cos 2k_y) $, 
$\epsilon^\perp_k = t^\perp_k \delta - 3J_\perp x_s^\perp/8$, and  
$\mu$ is the chemical potential. 
Free energy can be calculated using $E^{(\alpha\beta)}_k$,
\begin{eqnarray}
\displaystyle 
F_{MF}=E_0-2T\sum_{\alpha,\beta=\pm, k}
\log [1+\exp\big(-E^{(\alpha\beta)}_k /T\big)] + 2\mu N(1- \delta), 
\end{eqnarray}
where summation on $k$ is taken over the region $|k_x| + |k_y| \leq \pi$, and 
$E_0 = N [\frac{3J}{2}(x_s^2+y_s^2) + \frac{3J_\perp}{8} (x_s^\perp)^2]$ 
with $N$ being the total number of lattice sites within a plane.  
Self-consistency equations for the OPs and the chemical potential 
can be obtained by varying the free energy $F_{MF}$,\cite{Hamada,KK1} 
and we solve them numerically.  

$T_{FL}$ corresponding to the YBCO system is shown in Fig. 1. 
Here, the band parameters are chosen after Ref. 24;  
$t/J=2.5$, $t'/t=-0.3$, $t''/t=0.15$, $t^\perp_0/t=0.15$, and $J_\perp/J=0.1$.  
These parameters were chosen to reproduce experimental results for 
YBCO.\cite{Yamase} 
It is seen that the $T_{FL}$ for the type B flux phase is higher than that of 
the type A flux phase for $\delta \lesssim 0.15$.
The critical doping rate for the type 
A (B) flux phase is $\delta_c \sim 0.190$ (0.152).
Thus, $\delta_c$ in the bilayer model is consistent with that 
obtained in the experiment.\cite{Coving} 
At high doping rates, the $T_{FL}$ for the type B flux phase 
shows a reentrant behavior
at a low $T$ as in the case of the single-layer model.
This is because the nesting condition for the Fermi surface is changed 
for a large $\delta$, and then the incommensurate flux order, 
which is not taken into account in the present work, will be more favorable. 
For comparison, we also calculate the SC transition temperature $T_C$, 
using the self-consistency equations Eqs. (12)-(14) in Ref. 24. 
As seen, $T_C$ is always higher than $T_{FL}$ at any finite $\delta$, 
so that the stable solution in the bulk is the SC state. 

For comparison, we present the results for $t^\perp_0 = J_\perp =0$ in Fig. 2.   
$T_{FL1}$ ($T_{FL2}$) is that for $t/J=2.5$ and $t'=t''=0$ 
($t/J=2.5$, $t'/t=-0.3$, and $t''/t = 0.15$), 
and the corresponding SC transition temperature $T_{C1}$ ($T_{C2}$) is also shown. 
It is seen that $\delta_c$ is larger than that 
in Ref. 11, {\it i.e.},  $\delta_c\sim 0.11$ (0.08) 
for $t/J=4$ and $t'=t''=0$ ($t/J=4$, $t'/t=-1/5$, and $t'/t=1/6$, 
corresponding to the YBCO-type Fermi surface). 
This means that the larger $J/t$  is mainly responsible for the larger $\delta_c$,  
although the bilayer couplings (and also $t'$ and $t''$) may also affect it.

\begin{figure}
\begin{center}
\includegraphics[width=5.0cm,clip]{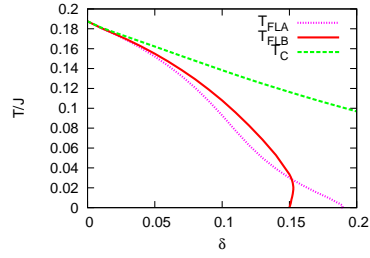}
\caption{(Color online) Bare transition temperature of the flux phase
of Type A ($T_{FLA}$), type B  ($T_{FLB}$), and superconductivity ($T_{C}$). 
See text for details. 
	 }
\end{center}
\end{figure}
\begin{figure}
\begin{center}
\includegraphics[width=5.0cm,clip]{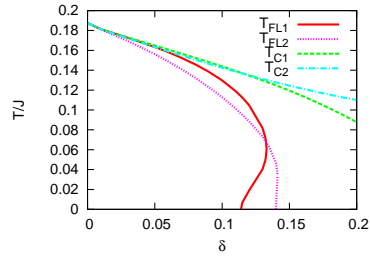}
\caption{(Color online) Bare transition temperature of the flux phase, $T_{FL}$,
and superconductivity, $T_{C}$, without interlayer couplings. 
 See text for details. 
	 }
\end{center}
\end{figure}

Near a  (110) surface, the $d_{x^2-y^2}$-wave SC order is strongly 
suppressed, and the flux phase would occur as in the single-layer model,  
with currents flowing along the surface.\cite{KK2}
In the type B flux phase,  
the current on the different layers will flow in opposite directions, 
and the magnetic field generated by these currents would vanish macroscopically.  
This may explain why no magnetic field is observed in some experiments 
for  the (110) surface of YBCO.\cite{msr}

In the single-layer model, the doping range where the flux phase exists 
is larger in inhomogeneous systems than  in uniform systems,  
because the incommensurate order not taken into account in the latter is 
expected in the former.\cite{KK2} 
We can expect that it is also the case in bilayer systems.  
Whether the transition from type B to A surface flux states ({\it \i.e.}, appearance 
of the local magnetic field near the surface) indeed occurs with increasing 
$\delta$ will be examined by BdG calculations. 
The local density of states should also be investigated to determine whether 
the peak splitting of zero bias conductance without a macroscopic magnetic 
field may be possible. These problems will be studied separately.

\begin{acknowledgments}
The author thanks M. Hayashi and H. Yamase for useful discussions. 
This work was supported by JSPS KAKENHI Grant Number 24540392. 
\end{acknowledgments}


\end{document}